\documentclass[10pt,sigconf,letterpaper]{acmart}
\usepackage{graphicx}
\usepackage[utf8]{inputenc}

\renewcommand\footnotetextcopyrightpermission[1]{} 

\usepackage{fancyvrb}
\usepackage{booktabs} 
\usepackage{listings} 
\usepackage{verbatim}
\usepackage{array}
\usepackage[normalem]{ulem}
\usepackage{xcolor}
\usepackage{color}
\usepackage{listings}
\usepackage[normalem]{ulem}
\useunder{\uline}{\ul}{}

\hypersetup{draft} 

\definecolor{todocolor}{rgb}{0.9,0.1,0.1}

\definecolor{mygreen}{HTML}{b7e1cd}
\definecolor{myred}{HTML}{f4c7c3}

\definecolor{stringColor}{rgb}{0.558215, 0.000000, 0.135316}
\lstdefinelanguage{JavaScript}{
  keywords={typeof, new, true, false, catch, function, return, null, catch,
  switch, var, if, in, while, do, else, case, break, const},
  keywordstyle=\color{blue}\bfseries,
  ndkeywords={class, export, boolean, throw, implements, import, this},
  ndkeywordstyle=\color{darkgray}\bfseries,
  identifierstyle=\color{black},
  sensitive=false,
  comment=[l]{//},
  morecomment=[s]{/*}{*/},
  commentstyle=\color{purple}\ttfamily,
  stringstyle=\color{stringColor}\ttfamily,
  morestring=[b]',
  morestring=[b]",
  numbers=left
}

\newcommand{\navigator}{}
\def\navigator/{\texttt{navigator}}

\begin{document}
\title{Characterizing Browser Fingerprinting \\ and its Mitigations}

\author{Alisha Ukani}

\begin{abstract}
    People are becoming increasingly concerned with their online privacy, especially
with how advertising companies track them across websites (a practice called
cross-site tracking), as reconstructing a user's browser history can reveal
sensitive information. Recent legislation like the General Data Protection
Regulation (GDPR) and the California Consumer Privacy Act have tried to limit
the extent to which third parties perform cross-site tracking, and browsers have
also made tracking more difficult by deprecating the most-common tracking
mechanism: third-party cookies. However, online advertising companies continue
to track users through other mechanisms that do not rely on cookies. This work
explores one of these tracking techniques: browser fingerprinting. We detail how
browser fingerprinting works, how prevalent it is, and what defenses can
mitigate it.
\end{abstract}

\maketitle
\pagestyle{plain} 

% ...

\raggedbottom

\section{Introduction}\label{sec:intro}

The Internet has quickly become part of every day life. We have come to rely on
it for entertainment, shopping, connecting with friends and family, and more.
However, this benefit comes with a hidden cost: tracking libraries are embedded
on many websites we visit, learning about our interests in order to show
targeted ads.

The use of tracking for advertising poses several privacy concerns. Users may
unintentionally reveal sensitive information through their web browsing, and
this sensitive information can be used against them. For example, privacy
advocates wish to prevent a situation where a user's insurance provider finds
out that they are looking up symptoms on WebMD, and as a result charges the user
a higher premium. These privacy risks are not hypothetical; famously, in 2012
Target mailed one household coupons for nursery and baby products -- identifying
the daughter as pregnant before her father found out~\cite{duhigg2012companies}.

Traditionally, trackers have used cookies to identify us on may of the websites
we visit. However, support for third-party cookies is being phased out by major
browsers. As a result, trackers are switching to techniques that do not rely on
cookies, such as browser fingerprinting.

In this work, we present an analysis of browser fingerprinting and how it works,
as well as the defenses that can mitigate it. The paper is structured as
follows. We start by introducing fingerprinting and its role in the ad ecosystem
in Section~\ref{sec:bkgd}. We then introduce core concepts of entropy and
k-anonymity in Section~\ref{sec:entropy}, which are used to measure how
effective fingerprinting is. Next we present a brief history of fingerprinting
and some common fingerprinting attributes in Sections~\ref{sec:origins}
and~\ref{sec:attr}. We then discuss its prevalence on the web in
Section~\ref{sec:adoption}. In Sections~\ref{sec:antiabuse}
and~\ref{sec:cookierespawn}, we present a positive and negative use case for
fingerprinting respectively. Finally, we compare the various defenses that
mitigate fingerprinting in Section~\ref{sec:defenses} and present takeaways in
Section~\ref{sec:discussion}.

\section{Background}\label{sec:bkgd}

This section provides context on the online tracking ecosystem and the role of
browser fingeprinting in this ecosystem.

\subsection{Ad Ecosystem}

Online advertising is a massive industry; one report found that online
advertising revenue in the US surpassed \$150B in 2021~\cite{iab2022outlook}.
Digital advertising relies on real-time bidding (RTB) markets, in which
advertisers bid for ad slots on websites (known as
publishers)~\cite{yuan2014survey}.

Advertisers are more likely to bid for ads shown to users in their target
demographic. Advertisers learn this information through data management
platforms (DMPs), which act as a middleman in this RTB market. DMPs track users
across websites to learn about their interests, construct demographic
information, and then sell this information in the RTB market.

Traditionally, DMPs use third-party cookies to track users across websites.
These DMP tracking scripts are embedded on many websites (making them third
parties), typically through \texttt{iframes}. When a user visits a website the
script is embedded on, the script will set a cookie in the user's browser with a
user identifier (UID). That cookie can be read by the same DMP script,
regardless of which site the script is embedded on. As that user continues
to visit other websites the same script is embedded on, the script can read the
value of the original cookie (i.e. the UID), and thus connect the user's current
session to the user's browsing history. In the rest of this work, we will refer
to DMPs as ``trackers.''

\subsection{Development of Cookieless Tracking}

In the past decade, public scrutiny and government regulation of online privacy
have increased dramatically. In particular, the 2016 General Data Protection
Regulation (GDPR) law in the E.U. ruled that people can be uniquely identified
through cookies, and so cookies count as personal data.

Though GDPR does not limit the collection or use of third-party cookies, it has
forced websites to be more transparent about how they collect information from
users for online advertising. This contributed to multiple browsers
announcing their intent to deprecate third-party
cookies~\cite{chrome3pc,firefox3pc,safari3pc}.

Now that trackers cannot easily collect information about users through
third-party cookies, they will switch to using cookieless tracking. Some forms
of cookieless tracking include navigational tracking, in which outgoing links
from websites redirect through trackers; CNAME cloaking, in which websites map a
subdomain to a tracker using DNS CNAME records, allowing them to share
first-party cookies; and browser fingerprinting, which we introduce in the next
section. Though cookieless tracking predates the deprecation of third-party
cookies, it has become increasingly important to understand and block in the
post-cookie web.

\subsection{Browser Fingerprinting}

Browser fingerprinting is a method of uniquely identifying users across websites
by querying information about the user's device (both the browser and the device
hardware). While any individual piece of information -- such as the fonts
installed on the device -- may not reveal unique information, once enough
properties are collected, a website may be able to uniquely identify a user
among millions of visitors to the site.

The research community has divided fingerprinting into two categories depending
on the source of the information on the user's device. \emph{Passive
fingerprinting} refers to a website collecting information from HTTP headers. In
contrast, \emph{active fingerprinting} relies on querying JavaScript APIs to
collect information. We primarily focus on active fingerprinting because
there are hundreds of APIs, which allows websites to identify users with finer
granularity than if they just perform passive fingerprinting. In addition, there
are more opportunities for interventions to prevent active fingerprinting, since
it would be difficult to remove access to HTTP headers to prevent passive
fingerprinting. We also focus on fingerprinting desktop devices because the
majority of the literature on fingerprinting exclusively considers desktop devices.

\section{Entropy and k-Anonymity}\label{sec:entropy}


In order to measure the efficacy of fingerprinting, it is important to
understand the intuition behind entropy~\cite{eckersley2010unique}; we do not
aim to provide a formal definition. With fingerprinting, entropy is used to
describe the amount of information a given attribute will convey about the user.
An attribute with higher entropy has greater variation in its values, and thus
provides more information about users. For example, image rendering is a
fingerprinting attribute with high entropy, as the resulting image looks
different depending on hardware and OS characteristics (we describe image
rendering in more detail in Section \ref{sec:canvas}). Many papers, especially
those describing new fingerprinting attributes, estimate the number of bits of
entropy these attributes provide. But these estimates heavily depend on the
dataset of fingerprints collected, and thus are difficult to compare across
studies. As a result, in this work we eschew discussion of specific entropy
estimates, though we do talk about entropy at a high level.

A related concept to entropy is k-anonymity~\cite{sweeney2002achieving}, which
in this context is used to compare fingerprints (or the values of an individual
fingerprinting attribute) across users. If a group of users have an anonymity
set of k, then at least k users have the same fingerprint. If k is large, then
it is difficult to differentiate users with the same fingerprint. If k is 1,
then that means each user has a unique fingerprint. Online trackers hope to
collect fingerprints so that k is as small as possible.


\section{Origins of Fingerprinting}\label{sec:origins}

The concept of browser fingerprinting was introduced by Mayer in his 2009
undergraduate thesis as the idea of ``quirkiness'': each user's browsing
environment is customized (and thus ``quirky''), and these differences can be
used to deanonymize users~\cite{mayer2009any}. To test this hypothesis, Mayer
collected fingerprints from 1,328 visitors of a university technology blog. The
fingerprints consisted of the \texttt{navigator} object (paying special
attention to
\texttt{navigator.plugins} and \texttt{navigator.mimeTypes}) and the
\texttt{screen} object. The \navigator/ object contains information about the
browser, such as the browser's language and User-Agent string, as well as the
list of installed plugins and MIME types supported by the browser. The
\texttt{screen} object contains information about the viewport, such as the
width, height, and pixel resolution. To test if each fingerprint was unique,
Mayer created a ground-truth dataset by setting a cookie for each user; if a
user visited the site multiple times, Mayer would know that this is a repeat
visit and not multiple distinct visitors with the same fingerprint. Ultimately,
he found that 96.23\% of the users could be uniquely identified with this simple
fingerprint.

Mayer's work offered a promising proof of concept. The following year, Peter
Eckersley published a larger-scale study that collected two orders of magnitude
more fingerprints than Mayer, and used a broader set of
attributes~\cite{eckersley2010unique}. Eckersley collected fingerprints from
470,161 users who visited the Electronic Frontier Foundation's website for his
study, known as Panopticlick.\footnote{This website is still available at
\url{https://coveryourtracks.eff.org/}.} The fingerprints included User-Agent
strings, HTTP ACCEPT headers, if cookies were enabled, screen resolution,
timezone, browser plugins (including plugin versions and MIME types), system
fonts, and other features. Eckersley was able to uniquely
identify 83.6\% of users. The most identifying attributes (i.e. the attributes
with the highest entropy) in descending order were the list of plugins and fonts
installed, the User-Agent string, the HTTP ACCEPT headers, and the screen
resolution. Eckersley also studied the stability of fingerprints by adding a
cookie to identify repeat visitors; 37.4\% of these repeat visitors had a
different fingerprint. However, Eckersley introduced an algorithm to identify
the previous fingerprint for a given user (if the user had both Flash and Java
installed) that had 99.1\% accuracy.

Eckersley's Panopticlick study was the first paper to show the potential of
browser fingerprinting for identifying users, and paved the way for future
research on this topic. Many of the attributes he identified are still widely
used in fingerprinting scripts; for example, as we discuss in more detail in
Section \ref{sec:fonts}, almost all fingerprinting scripts probe for fonts in
the same manner Eckersley described. We reference this study throughout the rest
of this work.

Fingerprinting has been used in the wild as early as 2009 for anti-fraud
purposes~\cite{mills2009device}. The earliest research studies to measure
fingerprinting adoption were published in
2013~\cite{acar2013fpdetective,nikiforakis2013cookieless}. Though we discuss
these studies in more detail in Section~\ref{sec:adoption}, we note that
Nikiforakis et al. identified three large, commercial fingerprinting
libraries~\cite{nikiforakis2013cookieless}. Acar et al. detected fingerprinting
behaviors, instead of starting with specific libraries, and identified 13
libraries present on the Alexa top-1M websites~\cite{acar2013fpdetective}. These
early studies suggest that fingerprinting libraries were rapidly developed by
many companies, even if overall adoption of these libraries was low.

The last event we highlight in this brief history is the creation of the
open-source fingerprintjs
library\footnote{\url{https://github.com/fingerprintjs/fingerprintjs}} in
2012~\cite{pinto2022fingerprintjs}. This open-source library is immensely
popular, and research has suggested that many fingerprinting libraries have
adopted the same techniques as fingerprintjs~\cite{acar2014web}.

\section{Fingerprinting Attributes}\label{sec:attr}

This section describes some of the common fingerprinting attributes and,
when possible, provides how many websites perform fingerprinting with that
attribute. It does not aim to provide a comprehensive list, but rather is a
primer on some common or notable attributes.

\subsection{Navigator Object}

As mentioned in Section \ref{sec:origins}, the \navigator/ object contains
several properties with information about the user's browser. Each property has
a name and a corresponding value. For example, the value of the
\texttt{navigator.plugins} property is a list of plugins, and the value of the
\texttt{navigator.mimeTypes} property is a list of MIME types. Mayer used both
of these properties as fingerprinting attributes~\cite{mayer2009any}. Not only
did he fingerprint people using the contents of these values (e.g. the actual
plugins installed), but he also fingerprinted people based on the order of the
values (e.g. the order in which the plugins were enumerated).

Nikiforakis et al. further investigated the \navigator/ object, and found that
it contained lots of device-specific information beyond the \texttt{plugins} and
\texttt{mimeTypes} properties~\cite{nikiforakis2013cookieless}. While Mayer
investigated the order of the properties' values, Nikiforakis et al. found that
there is also entropy in the order of the properties when enumerated. For
example, when a library prints the \navigator/ object, some browsers will print
the \texttt{plugins} property before the \texttt{mimeTypes} property, and vice
versa. The order of the properties is specific to the browser family and
version, and sometimes even specific to the same browser version on different
operating systems. In addition, some browsers have unique \navigator/
properties; for example, the \texttt{mozBrightness} property is only present in
Firefox browsers.

The \navigator/ object has been widely adopted for fingerprinting. For example,
the fingerprintjs library uses it for the \texttt{plugins} property like Mayer
proposed, as well as for the other various properties like Nikiforakis et al.
proposed. Though many studies explicitly log calls to the \navigator/
object~\cite{acar2013fpdetective,lin2022phish,durey2021fp,englehardt2016online},
none have reported the numbers on adoption. However, it is widely assumed that
any fingerprinting script will access this object.

\subsection{Canvas Fingerprinting and WebGL}\label{sec:canvas}

Canvas fingerprinting was first introduced by Mowery and Shacham in
2012~\cite{mowery2012pixel}. The \texttt{canvas} element allows websites to
render images using JavaScript. While the \texttt{canvas} element only supports
2D image rendering, developers can use the WebGL API to render 3D images. In
this study, Mowery and Shacham requested 300 Mechanical Turk participants to
load a website containing four images: three \texttt{canvas} images containing
text in various fonts (this method of rendering text using different fonts is
sometimes referred to as canvas fonts fingerprinting), and one WebGL
image. They found that these images produce high variation, even for a
seemingly-universal font like Arial, which leaked the device's OS and browser
family.

\begin{figure}
    \includegraphics[]{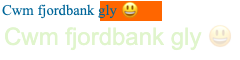}
    \caption{One of the canvas fingerprints generated by the fingerprintjs library. Each device's rendition will look slightly different.}
    \label{fig:canvas}
\end{figure}

Though the sample size of their participants was small, canvas fingerprinting
has become widely adopted. In 2014, Acar et al. found that 5.5\% of the Alexa
top 100K (5,542 sites) performed canvas fingerprinting, and that most of the
fingerprinting scripts seemed to copy the images rendered by
fingerprintjs~\cite{acar2014web}. We include an example of a \texttt{canvas}
image generated by fingerprintjs in Figure~\ref{fig:canvas}. A subsequent study
of the Alexa top 1M found that canvas fingerprinting was present on 14,371
websites~\cite{englehardt2016online}; we discuss this change in prevalence in
more detail in Section~\ref{sec:adoption}.

\subsection{Fonts}\label{sec:fonts}

Eckersley used the list of installed fonts as a fingerprinting attribute in the
Panopticlick paper, but collected fonts via Flash~\cite{eckersley2010unique}.
However, fonts can also be enumerated through JavaScript, which has become more
important for fingerprinting with the deprecation of Flash. Using JavaScript,
developers of fingerprinting libraries would hard-code a list of fonts, and for
each font the script will create a \texttt{span} element with some text in that
font. If the dimensions of the span (computed using the \texttt{offsetWidth} and
\texttt{offsetHeight} properties) are different from the dimensions of the span
in the system-default font, then the script detects the font is installed on the
user's machine.

In a 2013 analysis, Acar et al. measured fingerprinting adoption on the web
using font-probing as a proxy~\cite{acar2013fpdetective}. They found
that 404 of the Alexa top-1M websites loaded more than 30 fonts using
JavaScript, indicating fingerprinting. No other studies have highlighted the
frequency of font-based fingerprinting, but like the \navigator/ object, it is
often assumed to be used in most fingerprinting scripts. A more recent analysis
found that of 54 scripts that probe for fonts, the number of fonts probed ranges
from 66 to 594~\cite{durey2021fp}.

\subsection{Browser Extensions}\label{sec:attr:ext}

Browser extensions are add-ons created by external developers to enhance the
browsing experience for users, and many focus on enhancing user privacy. Despite
this goal, many have noted that these extensions can make users easier to
fingerprint. For example, Eckersley noted that some privacy-enhancing tools show
up in the User-Agent string, which can generate high
entropy~\cite{eckersley2010unique}. Some tools try to spoof the User-Agent
string to prevent users from being fingerprinted, but Nikiforakis et al. found
that the true browser and platform information can still be accessed via the
\navigator/ and \texttt{screen} objects~\cite{nikiforakis2013cookieless}. The
User-Agent string is also found in both HTTP headers and the \navigator/ object,
and some tools modified one but not both. So, if there is a discrepancy between
the User-Agent string and these JavaScript objects, a fingerprinting script can
detect the presence of this privacy-enhancing tool.

Even if an extension takes care to mitigate these discrepancies, it can still be
fingerprinted through its effects on the first-party website, as shown by Starov
and Nikiforakis~\cite{starov2017xhound}. This work introduced the XHound system,
which identifies the side-effects created by any browser extension. XHound
identified these side-effects by patching the extension code and generating
``honey pages,'' which are initially empty but dynamically add elements that are
expected by the extension. By running honey pages for the top-10K most popular
extensions, they found that 9.2\% produce detectable changes to the website's
HTML structure and thus can be fingerprinted. After discovering the unique
side-effect for each extension, the authors developed a proof of concept for how
trackers can identify the extensions a user has installed. The proof of concept
website focused on 30 extensions, and similar to XHound, created the HTML
elements expected by these extensions. The website then checked for the unique
side-effect of each extension. It is unknown if this method of fingerprinting
has been widely adopted.

\subsection{Battery Status}

The Battery Status API allows websites to learn a visitor's battery charge level
so that when a visitor's battery is low, the website can load fewer resources or
save changes more
frequently.\footnote{\url{https://developer.mozilla.org/en-US/docs/Web/API/Battery_Status_API}}
However, Englehardt and Narayanan confirmed that this API is used for
short-term fingerprinting~\cite{englehardt2016online}. They discovered multiple
scripts that query this API and combine it with other fingerprinting attributes,
like canvas fingerprints and the user's IP address.

\section{Fingerprinting Adoption}\label{sec:adoption}

\begin{table*}
    \small
    \begin{tabular}{|c|l|l|l|l|l|}
    \hline
     & \multicolumn{1}{c|}{Author} & \multicolumn{1}{c|}{\begin{tabular}[c]{@{}c@{}}Crawler\\ Technology\end{tabular}} & \multicolumn{1}{c|}{\begin{tabular}[c]{@{}c@{}}Fingerprinting\\ Detection\\ Procedure\end{tabular}} & \multicolumn{1}{c|}{\begin{tabular}[c]{@{}c@{}}Number of Sites\\ Analyzed\end{tabular}} & \multicolumn{1}{c|}{\begin{tabular}[c]{@{}c@{}}Prevalence of\\ Fingerprinting\end{tabular}} \\ \hline
    \begin{tabular}[c]{@{}c@{}}\textbf{FP-Inspector}~\cite{iqbal2021fingerprinting}\\ (2021)\end{tabular} & Iqbal et al. & OpenWPM (Firefox) & \begin{tabular}[c]{@{}l@{}}ML algorithm trained \\ on FP scripts known \\ to OpenWPM\end{tabular} & \begin{tabular}[c]{@{}l@{}}Alexa top 100K\\ (Successful: 71,112)\end{tabular} & 10.18\% (9,040 sites) \\ \hline
    \begin{tabular}[c]{@{}c@{}}\textbf{1M Site}\\\textbf{Analysis}~\cite{englehardt2016online}\\ (2016)\end{tabular} & \begin{tabular}[c]{@{}l@{}}Englehardt\\ and\\ Narayanan\end{tabular} & OpenWPM (Firefox) & \begin{tabular}[c]{@{}l@{}}Access to canvas, canvas\\ fonts, or WebRTC APIs\end{tabular} & \begin{tabular}[c]{@{}l@{}}Alexa top 1M\\ (Successful: 917,261)\end{tabular} & \begin{tabular}[c]{@{}l@{}}Canvas: 1.6\% (14,371)\\ Canvas fonts: <1\% (3,250)\\ WebRTC: <1\% (715)\end{tabular} \\ \hline
    \begin{tabular}[c]{@{}c@{}}\textbf{The Web Never}\\\textbf{Forgets}~\cite{acar2014web}\\ (2014)\end{tabular} & Acar et al. & Selenium (Firefox) & Canvas fingerprinting & Alexa top 100K & 5.5\% (5,542) \\ \hline
    \begin{tabular}[c]{@{}c@{}}\textbf{FP-Detective}~\cite{acar2013fpdetective}\\ (2013)\end{tabular} & Acar et al. & \begin{tabular}[c]{@{}l@{}}Selenium (Chromium),\\ CasperJS (PhantomJS)\end{tabular} & \begin{tabular}[c]{@{}l@{}}JS- or Flash-based\\ font enumeration\end{tabular} & \begin{tabular}[c]{@{}l@{}}JS detection: Alexa\\ top-1M home pages.\\ Also, for 100K sites,\\ visit 25 links/page\\ Flash: Alexa top-10K\\ home pages\end{tabular} & \begin{tabular}[c]{@{}l@{}}JS: 0.04\% (404)\\ Flash: 0.97\% (97)\end{tabular} \\ \hline
    \begin{tabular}[c]{@{}c@{}}\textbf{Cookieless}\\\textbf{Monster}~\cite{nikiforakis2013cookieless}\\ (2013)\end{tabular} & \begin{tabular}[c]{@{}l@{}}Nikiforakis\\ et al.\end{tabular} & Unknown & \begin{tabular}[c]{@{}l@{}}Inclusion of\\ fingerprinting scripts\\ from three\\ commercial libraries\end{tabular} & \begin{tabular}[c]{@{}l@{}}Alexa top 10K,\\ 20 pages per site\end{tabular} & 0.4\% (40) \\ \hline
    \end{tabular}
    \caption{Comparison of the academic papers that measure fingerprinting adoption on the web.}
    \label{tab:adoption}
    \end{table*}

This section compares the papers that measure how many websites fingerprint
their users. A high-level comparison is presented in Table~\ref{tab:adoption}.

The earliest papers to measure fingerprinting adoption were both published in
2013. In the ``Cookieless Monster'' paper, Nikiforakis et al. analyzed the
fingerprinting attributes used by three commercial fingerprinting
libraries~\cite{nikiforakis2013cookieless}. Then, they used a web crawler to
visit the Alexa top-10K sites, visiting up to 20 pages on each site, and
measured how many websites included one of these libraries (either through
\texttt{script} tags or through iframes). They found that 40 websites include at
least one of the three libraries, including Skype, which was the most popular
fingerprinting site. They also created a list of 3,804 domains that request
fingerprinting scripts, which do not have any relation to the Alexa rankings.
The majority of these websites were classified as spam by TrendMicro and
McAfee.

Nikiforakis et al. presented a thorough analysis of three
fingerprinting libraries, but was unable to identify new libraries. In contrast,
the ``FP-Detective'' paper by Acar et al. measured how many websites perform
fingerprinting regardless of which library they use~\cite{acar2013fpdetective}.
They classified a website as fingerprinting if it had a Flash object that
enumerates system fonts, or if it contained JavaScript that checks at least 30
fonts. They built a web crawler that could visit sites using Chromium or
PhantomJS (a lightweight rendering service based on WebKit). They crawled the
Alexa top 10K in Chromium and decompiled all Flash files to find that 97
websites performed Flash-based fingerprinting. To detect JS-based
fingerprinting, they used PhantomJS to crawl the Alexa top-1M home pages, and
also visited up to 25 pages from the landing pages of 100K websites. They found
that 404 websites perform fingerprinting using 13 different libraries, including
one library analyzed by Nikiforakis et al.~\cite{nikiforakis2013cookieless}.

While Acar et al. found a lower rate of JS-based fingerprinting than Nikiforakis
et al., they were able to find a greater number of fingerprinting libraries and
thus highlight the proliferation of fingerprinting libraries. However, one
limitation of this work is the use of PhantomJS, which does not fully emulate
browsers because it cannot load Flash or HTML5 objects. While PhantomJS is
well-suited for lightweight JavaScript collection, other papers cannot extend
this methodology to measure other fingerprinting attributes.

The next year, Acar published another fingerprinting study, this time focused on
canvas fingerprinting~\cite{acar2014web}. Acar et al. crawled the Alexa top 100K
and logged calls to various \texttt{canvas} functions. They found that 5.5\% of
websites performed canvas fingerprinting, with 95\% of these sites using the
same library from addthis.com. They also found that less popular sites were more
likely to fingerprint.

In 2016, Englehardt and Narayanan performed the largest study of fingerprinting
to date~\cite{englehardt2016online}. They developed the OpenWPM crawler, which
modifies Selenium to perform privacy measurements like logging JavaScript APIs
calls, or launching browsers with a seed profile. This platform has been used by
many subsequent
studies~\cite{fouad2022my,iqbal2021fingerprinting,englehardt2015cookies,das2018web,acar2020no}.
In this case, Englehardt and Narayanan used OpenWPM to measure how many of the
Alexa top-1M websites perform fingerprinting via \texttt{canvas} images,
\texttt{canvas} fonts, and WebRTC (an API to access media devices like webcams
and microphones). They found that \texttt{canvas} fonts fingerprinting was the
most common on popular sites; 2.5\% of the Alexa top-1K sites performed
\texttt{canvas} fonts fingerprinting, and the majority of these sites embedded
the same third-party: mathtag.com. However, \texttt{canvas} fingerprinting was
the most common overall, found on 1.6\% of sites. WebRTC was only found on 715
sites, making it very rare in comparison to canvas fingerprinting. Finally, the
authors found instances of websites performing fingerprinting using the
AudioContext and Battery Status APIs, both of which were not previously known to
be used for fingerprinting.

Interestingly, Englehardt and Narayanan found lower rates of canvas
fingerprinting than the prior work by Acar et
al.~\cite{acar2014web,englehardt2016online}. Englehardt and Narayanan
hypothesized that this decrease was due to the public backlash against trackers
triggered by the work of Acar et al. in 2014~\cite{acar2014web}.

The studies so far used heuristics to identify fingerprinting. In the
FP-Inspector system, Iqbal et al. trained a machine learning model with known
fingerprinting libraries to identify fingerprinting behaviors and
patterns~\cite{iqbal2021fingerprinting}. They found that their model had 99.9\%
accuracy and could detect 26\% more scripts than systems that used heuristics.
Then, they crawled the Alexa top 100K using OpenWPM and found that 10.18\% of
sites perform fingerprinting. This rate increased when looking at popular sites
(30\% of the Alexa top 1K) and news websites (14\%).

Because Iqbal et al. used machine learning instead of heuristics, they were able
to identify a broader set of fingerprinting libraries; they were not limited to
measuring specific attributes like the previous studies. This also let them
identify scripts that obfuscated their behavior to bypass existing heuristics.
For example, one script called \texttt{canvas} functions that have not been
identified by prior work. They also identified obfuscation where some scripts
perform fingerprinting in a ``dormant'' section, meaning the fingerprinting code
would only be executed when another function or event was triggered.

\textbf{Takeaways:}
FP-Inspector should be considered the final say on the general adoption of
browser fingerprinting because it detected new fingerprinting behaviors
instead of relying on heuristics. However, there is room for future work on more
targeted measurement studies. For example, Section~\ref{sec:antiabuse:adoption}
details the need for a large-scale measurement study of fingerprinting on
authentication pages.

But the web is constantly evolving, and there are specific events that would
motivate further measurement studies on fingerprinting adoption. For example,
once Chrome deprecates third-party cookies, a measurement study could evaluate
whether advertisers completely switch to tracking users via fingerprinting. Or,
if browsers introduce any further fingerprinting protections, a measurement
study could evaluate how effective those protections are.







\section{Fingerprinting for Anti-Abuse and Authentication}\label{sec:antiabuse}

As mentioned in Section~\ref{sec:origins}, the first websites to adopt
fingerprinting claimed it was for anti-fraud and abuse purposes, not for
tracking. However, it is difficult to identify the purpose of fingerprinting on
any given website -- fingerprinting can be used for anti-abuse, tracking, or
both. In this section, we will discuss papers that exclusively focus on
fingerprinting for anti-abuse and authentication purposes.

We focus on three attack scenarios, as described by Durey et
al.~\cite{durey2021fp}. The first scenario is stolen credentials, in which a
user's login information is acquired by an attacker. Fingerprinting can be
useful here to detect when the attacker attempts to log-in, and the website can
then raise a multi-factor authentication (MFA) prompt. Similarly, cookie
hijacking can allow an attacker to access user accounts, and fingerprinting can
detect when a cookie is used on a different device. Lastly, fingerprinting can
be used for bot detection, which can mitigate harms like DDoS attacks and click
fraud.

\subsection{Criteria for Authentication}

Many researchers have proposed using fingerprinting for these anti-abuse and
authentication
purposes~\cite{andriamilanto2021large,rudametkin2021improving,goethem2016accelerometer,preuveneers2015smartauth,unger2013shpf,alaca2016device}.
One of the largest empirical studies of fingerprints proposed that fingerprints
are well-suited for these purposes because they are unique, stable, and easy to
collect~\cite{andriamilanto2021large}. This study defined a set of criteria for
``biometrics'' to be usable and practical for authentication: the biometric
should be universally present, distinct for each person, stable (i.e. not
changing over time), and able to be collected. Furthermore, they say that a
biometric authentication scheme is practical if it is performant, robust against
attacks, and if users accept it in their daily lives. The authors specifically
focus on the distinctiveness, stability, and performance of fingerprints to
satisfy the criteria.

In this study, Andriamilanto et al. collected over 4M empirical fingerprints,
collecting 216 attributes (and deriving a further 42 attributes for a total of
262). They found that 81\% of the fingerprints were unique, satisfying the
distinctiveness property. For the performance property, the authors noted that
most of their chosen attributes are HTTP headers or JavaScript properties, which
can be collected in less than 5ms. The 33 attributes that took longer than 5ms
to collect included detection of browser extensions, measuring media elements,
and detection of browser components, e.g. the list of speech synthesis voices.
Attributes also have a low memory footprint, with 137 attributes having a size
less than 5 bytes each. Only 20 attributes required more than 100 bytes, such as
listing the properties of the \texttt{navigator} object. Overall, the authors
claim that these latencies and memory sizes are low enough to make fingerprints
performant.

To measure fingerprint stability, users were assigned a six-month cookie
containing a UID. The authors found that 90\% of the fingerprint attributes are
the same after six months. Only six attributes were highly
unstable.\footnote{The authors measured stability using \textbf{sameness rate},
defined as ``the proportion of the consecutive fingerprints where the value of
the attribute stays identical.''~\cite{andriamilanto2021large}} These unstable
attributes include the width and height of newly-created \texttt{div} elements
-- a new fingerprinting attribute the authors identified and collected -- as
well as an HTTP header for setting cache policies and WebRTC. While WebRTC APIs
have been found in fingerprinting scripts~\cite{englehardt2016online}, the
authors found them to be unstable because of the API's low adoption across
websites, and its use of local IP addresses~\cite{andriamilanto2021large}.

Ultimately, the authors found that browser fingerprinting satisfied their
distinctiveness, stability, and performance criteria, and thus can be used for
authentication.

\subsection{Potential Attacks}

Andriamilanto et al. proposed using fingerprinting for authentication, but also
warned that this authentication system is not
foolproof~\cite{andriamilanto2021large}. Similar to various password attacks, an
attacker can try to impersonate a victim's fingerprint using brute force,
dictionary attacks, replay attacks, or relay attacks. In the latter two attacks,
the attacker tricks the user into visiting an attacker-controlled site that
collects the victim's fingerprint. The attacker can then spoof their own
fingerprint to match the victim.

This attack scenario was proposed and empirically evaluated by Lin et
al.~\cite{lin2022phish}. Specifically, this work hypothesized that after a user
logs into a website using a MFA prompt and clicks ``remember me,'' the website
uses browser fingerprinting to decide if subsequent MFA prompts should be shown.
Then, an attacker can use a phishing site with a fingerprinting script to
collect victim credentials and bypass MFA prompts.

To evaluate this attack, the authors built two browser extensions, one to
collect fingerprints and one to spoof them. The former collected the properties
of the \texttt{navigator}, \texttt{window}, and \texttt{screen} objects,
recorded the result of any images rendered using \texttt{canvas} or WebGL, and
logged information about any font properties accessed by the website. As a
result, the authors were able to detect fingerprinting via JavaScript objects
(\navigator/, \texttt{window}, and \texttt{screen}), canvas and WebGL, font
enumeration, timezone, WebRTC, and AudioContext. The authors also developed a
spoofing extension, which modified the browser's JavaScript objects and replaced
font information and \texttt{canvas}/WebGL images with the stored values. 

The authors manually tested 300 websites to see if they were vulnerable to the
proposed attack. They logged into an account from the primary (victim) device
and clicked ``remember this device.'' They logged out, logged back in, and
confirmed that no MFA prompt is shown. Then, they logged into the account from a
secondary (attacker) device and confirmed that a MFA prompt was shown. Finally,
they used the spoofing extension to spoof the primary device's fingerprint on
the secondary device, and then logged in and checked if a MFA prompt was shown.
They found that 14 websites use fingerprinting to remember devices, and that
nine of these were vulnerable. In some of these cases, spoofing the fingerprint
also suppressed email notifications to the user about new device logins. The
five robust websites checked for matching IP addresses. An easy solution would
be for websites to check if the IP address from a new login attempt matches the
IP address used for previous logins; however, if users start to rotate IP
addresses frequently (e.g. due to Apple iCloud Private Relay), it is unclear if
websites will stop trying to match IP addresses.

Lastly, Lin et al. analyzed more than 153,000 actual phishing websites from a
two-year span to see how many of these sites perform fingerprinting. They
performed dynamic analysis and analyzed the JavaScript APIs called by the
website, finding that the majority of phishing sites fingerprinted users. They
also found that a smaller fraction of these phishing sites collected enough
fingerprinting attributes to spoof a fingerprint for the target site, thus
creating the capability to bypass MFA prompts.

\subsection{Adoption}\label{sec:antiabuse:adoption}

Lin et al. also performed a small-scale measurement study of fingerprinting on
login pages on popular websites~\cite{lin2022phish}. They crawled the Alexa
top 20K and detected login pages for 11.5k of the sites using a set of
heuristics developed in prior work~\cite{drakonakis2020cookie}. They found the
majority of websites fingerprinted visitors of their login pages, with
higher-ranked websites using more advanced fingerprinting techniques (canvas
fonts, WebRTC, and AudioContext) on their login pages compared to their home
pages.

Another study by Durey et al. investigated fingerprinting on various internal
pages for a smaller set of websites~\cite{durey2021fp}. This study analyzed
1,485 pages from 446 websites, including a variety of page categories: landing,
sign-up, sign-in, payment, shopping cart, and content (i.e. any other internal)
pages. They find that while 23\% of the home pages performed fingerprinting,
that rate increased for shopping cart (33.8\%) and sign-up (31.1\%)
pages. They also identified 14 fingerprinting scripts from security companies,
which seemed tailored for payment platforms, fraud prevention, and bot
detection.

\textbf{Takeaways:}
To the best of our knowledge, these studies are the only ones that address
fingerprinting adoption on authentication pages. They indicate that
fingerprining is happening, but leave room for a more sophisticated analysis
performed on a larger set of websites. A key challenge here is automatically
identifying the login and sign-up pages for a set of websites; Durey et al. use
manual analysis to find these pages~\cite{durey2021fp}, while the heuristics Lin
et al. use are simple and rely on regex strings with English phrases (and thus
do not work well for websites in other
languages)~\cite{lin2022phish,drakonakis2020cookie}.

\section{Cookie Respawning}\label{sec:cookierespawn}

A major privacy concern for fingerprinting is that it can be used for
cookie respawning. In cookie respawning, a user visits a website that
sets a cookie for the user based on their fingerprint. The user then clears
their cookies, and later visits the website again. The website is able to
regenerate the original cookie (because the user has the same fingerprint) and
store it in the user's browser. Peter Eckersley predicted that fingerprinting
could be used for this purpose in 2010~\cite{eckersley2010unique}. Cookie
respawning is concerning because it can be used to deliberately ignore user
actions; even if a privacy-conscious user regularly clears their cookies,
websites can restore those cookies on the user's next visit.

Another harm of cookie respawning, described by Acar et al., is that if a user
enables third-party cookies, then cookie respawning can help advertisers track
users over longer periods of time~\cite{acar2014web}. If a user enables
third-party cookies, then a third-party script can
track users who visit any site that script is embedded on. So, the script can
track a user across websites until the user clears their cookies; at that point,
the script can start tracking the user again, but cannot link this new set
of websites to the user's previous browsing history. However, with cookie
respawning, the script is able to associate the user with the websites they
visited before clearing cookies. So, cookie respawning gives advertisers access
to user data for longer periods of times.

Several studies have found that some websites use Flash cookies to regenerate
HTTP cookies that users have removed~\cite{acar2014web, soltani2010flash,
ayenson2011flash}. Similar to regular cookies, Flash cookies (also known as
local shared objects) are text files stored in the user's browser. Flash cookies
can only be set by websites hosting Adobe Flash content. In 2014,
Acar et al. found that 30 websites of the Alexa top 10K respawned HTTP cookies
from Flash cookies~\cite{acar2014web}.

Eight years later, cookie respawning is still
occurring through browser fingerprinting. Recently, Fouad et al. conducted a
study to measure the prevalence of this respawning~\cite{fouad2022my}. They set
up a web crawler on two different machines (Machines A and B) that presented
different fingerprints. They visited the Alexa top 30K on each machine and
collected the cookies stored by each website. Then, they cleared the cookies on
Machine A and repeated the crawl on Machine A. They identified respawned cookies
as the cookies that appeared on both crawls on Machine A, excluding any of these
cookies that appeared on the Machine B crawl (as those cookies are not
user-specific cookies).

With this methodology, Fouad et al. found 5,144 user-specific cookies
respawned across 4,093 websites. They further filtered this set of cookies to
the cookies respawned using a set of eight fingerprinting features: the HTTP
ACCEPT language header, geolocation, User-Agent string, Do Not Track header,
WebGL, Canvas, IP address, and time zone. They found 1,425 cookies across 1,150
websites were respawned using at least one of these features. The most common
feature used to respawn cookies was the machine's IP address, used for 672
cookies. This finding suggests that if a user's IP address changes frequently
(e.g. for users of Apple's iCloud Private Relay), then websites will be unable
to respawn cookies as easily.

Finally, Fouad et al. analyzed the legality of cookie respawning and found that
there is no clear legal interpretation from GDPR and its predecessor the
ePrivacy Directive (ePD). However, they noted that the practice of cookie
respawning violates the GDPR principle that personal data should be processed
fairly. In addition, the privacy policies of the top-10 most popular websites
that respawned cookies did not mention this practice; this violates the GDPR
principle that personal data processing must be transparent to users.

\section{Defenses}\label{sec:defenses}

In this section we introduce four main categories of defenses: reducing entropy,
increasing entropy, blocking, and increasing visibility. Reducing entropy
removes information about users so it is harder to uniquely identify them.
Increasing entropy adds randomness to JavaScript APIs so that there are multiple
fingerprints per user, which prevents multiple visits from being linked.
Next, blocking can refer to either blocking trackers or blocking access to APIs
to prevent fingerprinting from occurring in the first place. Finally, increasing
visibility allows fingerprinting to occur, but makes it visible to
users. Though our work introduces this framework for analyzing defenses, we will
define subcategories that have already been introduced in the
literature~\cite{al2020too,laperdrix2020browser}.

\subsection{Reducing Entropy}

Reducing entropy limits the information that fingerprinting scripts are able to
collect, which makes it harder to uniquely identify devices. There are two main
subcategories: reducing the amount of information returned by JavaScript APIs,
and presenting uniform fingerprints.

\begin{figure}
\begin{lstlisting}[caption={An example User-Agent string.}, label={lst:ua1}]
Mozilla/5.0 (Linux; Android !\colorbox{myred}{12; Pixel 5}!) AppleWebKit/537.36 (KHTML, like Gecko) Chrome/95.!\colorbox{myred}{0.4638.16}! Mobile Safari/537.36
\end{lstlisting}
\end{figure}

\begin{figure}
\begin{lstlisting}[caption={The User-Agent string from Listing \ref{lst:ua1} after Chrome's User-Agent reduction feature.}, label={lst:ua2}]
Mozilla/5.0 (Linux; Android !\colorbox{mygreen}{10; K}!) AppleWebKit/537.36 (KHTML, like Gecko) Chrome/95.!\colorbox{mygreen}{0.0.0}! Mobile Safari/537.36
\end{lstlisting}
\end{figure}

\subsubsection{Reducing JavaScript API information}

Many JavaScript APIs return precise, detailed information. For example, a
User-Agent string is shown in Listing \ref{lst:ua1}. This User-Agent string
indicates that the device is a Pixel 5 phone running Android 12 and Chrome
version 95.0.4638.16. This level of detail -- specific phone models and minor
browser version numbers -- creates more entropy and thus produces smaller
anonymity sets.

The goal of this defense category is to reduce the amount of information
returned by APIs so they contain less granular information. Eckersley noted that
sometimes the granular information, like specific browser versions, is helpful
for debugging~\cite{eckersley2010unique}. If a website developer sees many
errors associated with a specific browser version, this may indicate that some
change in the browser is causing the errors. In these cases, browsers must weigh
the trade-offs between helping website developers and preserving user
privacy. However, Eckersley also noted that some sources of entropy are
unnecessary; for example, Flash apps returned a list of fonts sorted in a
specific order, and this order could be used for fingerprinting. If this list of
fonts was sorted before being returned, then fingerprinting scripts would gain
less information about the user.

This defense is starting to be implemented by browsers, especially for the
User-Agent string. Google Chrome is already testing out simplified User-Agent
strings, which remove the OS version, device, and minor browser
version.\footnote{https://developer.chrome.com/docs/privacy-sandbox/user-agent/snippets/}
For example, the User-Agent string in Listing \ref{lst:ua1} would be transformed
to look like the one in Listing \ref{lst:ua2}. Websites can still request the
full information using a new UA Client Hints API; browsers can choose to answer
these requests upon certain conditions, like letting users decide which values
can be revealed to websites~\cite{uaclienthints}. Ultimately, the success of
this work to reduce fingerprinting will depend on how actively websites use the
Client Hints API, and to what extent browsers will freely answer these API
requests.

\subsubsection{Uniform fingerprints}

To take the idea of reducing entropy to the extreme, browsers can render
fingerprinting scripts useless by presenting the same fingerprint for every
user.

The most notable example of this defense is the Tor browser. Tor users present
the same fingerprint across 24 different attributes~\cite{laperdrix2020browser}.
For example, Tor removes plugins, sets a default set of fonts, and returns a
blank image for all \texttt{canvas}
functions~\cite{acar2014web,laperdrix2020browser}. Firefox also implements many
of these same defenses.\footnote{\url{https://wiki.mozilla.org/Security/Fingerprinting}}

Unfortunately, these protections are brittle and can impact the user's browsing
experience. For example, consider the \texttt{screen} object, which is used for
fingerprinting and contains information about the device's screen size and
resolution. Tor presents a uniform \texttt{screen} object by setting the browser
window size to be 1,000 by 1,000 pixels~\cite{laperdrix2020browser}. If the user
changes the window size, they can be identified as a Tor user with that specific
screen size -- which may be more uniquely identifying than if the user chose a
more popular browser.

This example hints at another shortcoming of this defense. Though presenting a
uniform fingerprint protects Tor users, it can still leave them vulnerable in
some ways because they are now easily identifiable as Tor users. This is similar
to a finding discussed in Section \ref{sec:attr:ext}: some techniques of
blocking fingerprinting actually leaves users more fingerprintable. Being
identified as a Tor user is still better than having a unique fingerprint across
all users. But this defense works best when a large number of people are
presenting the same fingerprint. So, this defense would be more effective if it
were adopted by other browsers.

\subsection{Increasing Entropy}\label{sec:defense:increase}


Instead of decreasing the set of possible values for JavaScript APIs, another
defense is to increase the set of possible values so much that each user has
multiple fingerprints. As a result, users will have different fingerprints for
different websites, and even different fingerprints for subsequent visits to the
same website, and therefore cannot be tracked effectively. The way this is
implemented is by returning random values as the output of various JavaScript
APIs (also called ``attribute spoofing'' by Al-Fannah et al.~\cite{al2020too}).

One tool that has implemented this defense is the (now defunct) Firefox
extension Firegloves~\cite{acar2013fpdetective}. Firegloves returned random
values for attributes like screen resolution; User-Agent properties like the
browser vendor, browser version, and platform; and the HTML properties
\texttt{offsetWidth} and \texttt{offsetHeight}. Unfortunately, Acar et al. found
several limitations that still enable fingerprinting. While not a concern today,
websites could use Flash to learn the browser's screen resolution. Furthermore,
websites could learn the true platform by checking \texttt{navigator.oscpu}, and
check if the browser is Firefox by checking for the presence of Firefox-specific
\navigator/ properties like \texttt{navigator.mozCameras}. Finally, websites
could learn the dimensions of HTML elements using the
\texttt{getBoundingClientRect()} method instead of using \texttt{offsetWidth}
and \texttt{offsetHeight}.

More recent work has increased entropy without creating discrepancies like
Firegloves. While there are several such studies, we focus on two papers:
FP-Block~\cite{torres2015fp} and
PriVaricator~\cite{nikiforakis2015privaricator}. Both of these papers were
published in 2015 and both focus on the idea of increasing entropy. However,
they take very different approaches in terms of the fingerprinting attributes
they defend against and the layer of the stack in which the defense is
implemented. As a result, they also provide interesting insights into how
fingerprinting defenses should be implemented.

\subsubsection{FP-Block}

FP-Block is a Firefox extension that randomizes values in five different
categories of fingerprinting attributes: browser information (e.g. the
User-Agent string), language, OS/CPU information, screen (e.g. width and
height), and timezone~\cite{torres2015fp}. For each new website the user visits,
FP-Block will start with the current fingerprint, and modify the values of at
least two attributes from at least two categories. When choosing new values,
FP-Block relies on a distribution of values from prior work; for example, if
FP-Block chooses the browser to be Chrome and is then deciding on what platform
the fingerprint should have, it will pick the Windows platform with 89\%
probability because 89\% of the studied Chrome users have Windows machines. To
actually spoof these values, FP-Block modifies the getter functions for the
\navigator/ and \texttt{screen} objects. In addition, FP-Block adds noise to
\texttt{canvas} images, but stores the rendered image so that subsequent visits
to the same website present the same fingerprint; FP-Block aims to prevent
cross-site tracking, but allows websites to identify repeat visits.

To evaluate FP-Block, the authors created a set of test websites that ran the
fingerprintjs library. They confirmed that different browsers visiting the same
site had different fingerprints, and that the same browser visiting different
tests websites also had different fingerprints. They also confirmed these
results for four other fingerprinting libraries, including BlueCava, which
Pri-Varicator also defends against~\cite{nikiforakis2015privaricator}. Finally,
they qualitatively confirmed that the extension did not crash frequently, caused
minimal breakage, and only slowed down websites that ran lots of JavaScript
code.

\subsubsection{PriVaricator}

PriVaricator used randomness to control access to installed fonts and plugins
while attempting to minimize website
breakage~\cite{nikiforakis2015privaricator}. They focused only on fonts and
plugins because these are fingerprinting attributes with high entropy, and
because canvas fingerprinting was not widely adopted at the time. They also
noted that modifying the User-Agent string can affect the website's HTML
output, which can cause breakage.

The authors randomized plugins with a parameter that specifies what percentage of
plugins to reveal; the authors did not attempt to list plugins that were not
actually installed by the user. For fonts, recall from
Section~\ref{sec:fonts} that a fingerprinting script will create an HTML
element with text rendered in the font the script is trying to probe. If the
font is not installed, it will be rendered in the default font. So, if the
element's \texttt{offsetWidth} and \texttt{offsetHeight} dimensions are
different than the dimensions of the default-font text, then the font is
installed. The authors increased entropy through three strategies: setting
\texttt{offsetWidth} and \texttt{offsetHeight} to zero, setting each one to a
random value between 0 and 100, and modifying the actual values with $\pm 5\%$
noise. Finally, the authors introduced two parameters for when to ``lie'' (i.e.
after how many \texttt{offsetWidth} and \texttt{offsetHeight} access to start
adding randomness) and what percentage of \texttt{offsetWidth} and
\texttt{offsetHeight} accesses to modify.

PriVaricator was implemented directly in Chromium. The authors found that
PriVaricator was effective at creating unique fingerprints for BlueCava,
fingerprintjs, Coinbase (which fingerprinted users in a ``pay with Bitcoin''
\texttt{iframe}), and a research prototype library. To test how much breakage
PriVaricator causes, they crawled the Alexa top 1K using 48 combinations of
parameters that resulted in unique fingerprints across all the services tested.
They found that the strategy of randomizing with noise resulted in the least
amount of breakage, only affecting 0.6\% of sites on average. The breakage
included not loading ads, error pages, and image carousels; however, of 100
manually-inspected websites marked as broken, 92 were false positives due to
dynamic content like pop-ups.

\subsubsection{Comparison of FP-Block and PriVaricator}

FP-Block and PriVaricator had the same key insight of using randomness to create
multiple fingerprints per user, but they implemented their defenses in very
different ways. FP-Block is a Firefox extension that spoofs values by modifying
getter functions; in contrast, the PriVaricator authors noted that
fingerprinting scripts can fetch these getter functions and check if they are
modified. For this reason, PriVaricator was implemented directly in Chromium and
is thus more robust to scripts trying to obfuscate their behavior.

While PriVaricator focused solely on fonts and plugins, FP-Block spoofed values
for 22 different fingerprinting attributes used by commercial trackers,
including fonts and plugins. Though the PriVaricator authors justified their
choice by arguing that modifying the User-Agent string can result in breakage,
FP-Block still modified values for many other attributes that are less likely to
cause breakage, such as the timezone and various HTTP header values.

However, it is difficult to quantify the breakage caused by FP-Block because the
authors did not present a quantitative evaluation. In contrast, PriVaricator had
a strong, formal evaluation of breakage and performance impact.

\textbf{Takeaways:}
Future defenses can take away some important lessons from these two papers.
First, both papers learned from the techniques of real, commercial
fingerprinting libraries; this focus helped ensure the defenses could protect
users from the libraries are most likely to encounter while browsing the web. In
addition, future defenses should weigh the trade-offs between writing browser
extensions (which are lightweight and easy to prototype) and modifying the
browser (which can prevent libraries from noticing the side-effects of an
extension). Finally, we recommend that future defenses consider a large set of
fingerprinting attributes like FP-Block did, and conduct a comprehensive
evaluation of site breakage and performance impact like PriVaricator did.

To conclude this section, we note that the defense of increasing entropy has
been implemented by the Brave browser. Brave randomizes \texttt{canvas} and
WebGL rendering, as well as the User-Agent string, web audio, plugins, and
several other
attributes.\footnote{https://github.com/brave/brave-browser/wiki/Fingerprinting-Protections}

\subsection{Blocking}

The previous two defense categories allowed fingerprinting scripts to run, but
prevented these scripts from collecting the information they need. In contrast,
the blocking category prevents these scripts from running in the first place, or
prevents the scripts from accessing the required APIs.

\subsubsection{Blocking Trackers}\label{sec:defense:block:tracker}

Browser extensions are the most common way to identify and block trackers on the
web. Some notable extensions are AdBlock Plus, which blocks ads and third-party
tracking scripts; Privacy Badger, which blocks third-party trackers observed to
be tracking the user on multiple sites; and Ghostery, which also blocks
third-party trackers based on a blocklist. In particular, Englehardt and
Narayanan found that Ghostery is effective at blocking popular third parties and
cookies with UIDs, but is limited by its use of blocklists; identifying and
blocking new trackers is a slow, and sometimes manual
effort~\cite{englehardt2016online}.

Despite this limitation, browsers are also starting to adopt this blocking
strategy. In particular, Firefox blocks fingerprinting scripts identified by the
Disconnect blocklist~\cite{englehardt2020firefox}.

\subsubsection{Blocking API Access}

Instead of blocking the fingerprinting scripts, browsers can block the APIs used
to perform fingerprinting. If applied to enough APIs, it can be effective at
stopping fingerprinting. However, there is a trade-off: deprecating APIs will
also prevent benign sites from using them to provide enhanced functionality.

One approach to determining which APIs to deprecate is to perform an audit on
how each API is being used. For example, Olejnik et al. studied the use of the
Battery Status API on the Alexa top-50K sites~\cite{olejnik2017battery}. They
found that the API was used on 841 sites, mostly by third parties. Of these 33
third parties, 16 were using the API for tracking. The benign usage was
attributed to YouTube, which used it for embedded video performance metrics, and
a performance measurement library. Both
Firefox\footnote{\url{https://bugzilla.mozilla.org/show_bug.cgi?id=1313580}} and
Brave\footnote{\url{https://github.com/brave/browser-laptop/issues/1885}} have
removed access to the Battery Status API because it is mostly used for
fingerprinting, though other browsers like Chrome and Edge still support
it.

Al-Fannah et al. have also proposed a more moderate approach, such as limiting
the functionality of some APIs~\cite{al2020too}. For example, they proposed
keeping the \texttt{canvas} API for creating images, but preventing websites
from retrieving the rendered images. They also proposed controlling access to
APIs based on context. This could mean making sure only certain websites could
trigger certain APIs (e.g. only video chat sites can use the WebRTC API), or
that third parties cannot access any API. Again, the primary limitation of this
approach is that it can prevent benign websites from accessing APIs. It is
unclear if browsers are willing to protect user privacy at the expense of a
richly interactive web.

\subsection{Increasing Visibility}

The last type of defense allows fingerprinting to occur as it is, but makes this
behavior visible to users in real time as they browse the web. One user study
found that users understand that fingerprinting is used for tracking purposes,
but have a poor understanding of its prevalence on the
web~\cite{pugliese2020long}. Some participants believed they would not be
fingerprinted if they visit ``trustworthy'' sites and avoid clicking on
``dangerous links,'' when in reality news websites are more likely to
fingerprint users than websites with illegal
content~\cite{iqbal2021fingerprinting}. This motivates the need for a way to
notify users in real time if they are visiting a site that is performing
fingerprinting.

Weinshel et al. designed an interface to notify users about which websites are
tracking them, including some forms of fingerprinting~\cite{weinshel2019oh}.
This study found that users were surprised by how often they were tracked and
were motivated to install tools that block tracking. This promising study
indicates that users want to know when they are tracked and that this
information will help them take action to preserve their privacy; however, the
browser extension developed in this work did not show any real-time indications
of tracking, and instead required users to click into the extension to get
information for every page they visit.

To the best of our knowledge, the only paper to describe a system that notifies
users of fingerprinting in real time was developed by Fietkau et
al.~\cite{fietkau2021elephant}. This work presented a browser extension whose
icon changes from green to yellow to red when extensive fingerprinting is
detected on a website. To detect fingerprinting, the authors first developed a
list of 40 JavaScript APIs commonly used for fingerprinting. At a high level,
the more attributes the website accessed, the higher the website's
fingerprinting score became. More concretely, the authors analyzed the
distribution of attributes accessed by the Alexa top 10K to determine thresholds
for ``low,'' ``medium,'' and ``high'' amounts of fingerprinting. Scores further
increased if the websites accessed attributes that are high-entropy and thus
more uniquely identifying.

\textbf{Takeaways:}
Fietkau et al.'s paper presented a very simple detection technique, especially
in comparison to more sophisticated analysis done by Iqbal et al. in the same
year~\cite{iqbal2021fingerprinting}. Furthermore, the authors did not conduct a
user study to evaluate their extension, so it is unclear if users would find
this extension helpful. We believe there is more work to be done in this area.



\section{Discussion and Conclusion}\label{sec:discussion}

We conclude this work by evaluating the effectiveness of each defense, and
exploring avenues for future work.

\subsection{Evaluation of Defenses}

There is no one technique that will instantly stop browser fingerprinting. Much
like security, there will always be an arms race between trackers developing new
fingerprinting attributes and privacy advocates attempting to identify and
defend against those new attributes. However, there are some defenses that can
help make fingerprinting less effective for tracking users.

Of the categories we defined, the least effective is making fingerprinting
visible to users. While there is important moral value in making this behavior
visible, this is not an effective solution for stopping fingerprinting. People
can quickly become desensitized to frequent warnings or alerts about online
tracking~\cite{burgess2018tyranny}. Furthermore, if users are constantly alerted
that websites are fingerprinting them, but there is no immediate action users
can take to limit this tracking, then the warnings can make users feel helpless
and pessimistic.

The reducing entropy defense category is helpful, but its effectiveness is
limited. Trying to present a uniform fingerprint for all users is brittle
because actions like changing the browser window size can introduce variation.
Reducing the entropy from individual APIs is more robust, but only works for
certain APIs. While the User-Agent string is the prime example for this
strategy, reducing entropy does not make sense for canvas and WebGL
fingerprinting, or for testing the presence of certain fonts.

A more promising defense category is increasing entropy. Unlike the previous
category, this category can be applied to many fingerprinting techniques.
Section~\ref{sec:defense:increase} highlighted systems that increase entropy for
canvas fingerprints, fonts, and plugins. In particular, PriVaricator caused
minimal breakage and performance overhead, making this defense more practical
and appealing for wider adoption~\cite{nikiforakis2015privaricator}. The best
endorsement of this defense is the fact that Brave has adopted it for a variety
of fingerprinting attributes. We hope that other browsers adopt this defense as
well.

The last defense category, blocking, has the most promise to stop fingerprinting
completely, but is limited by the arms race between trackers and blockers. As
shown in Section~\ref{sec:defense:block:tracker}, browser extensions that block
third-party trackers have been shown to be effective. They also provide a
stronger guarantee than the previous defenses; instead of making fingerprinting
weaker, they block fingerprinting scripts from running in the first place. However, trackers have
started obfuscating their code and combining it with benign functionality so
that blocking trackers also breaks the first-party website. Fortunately there
has been increased attention in separating tracking code from benign JavaScript
code in these libraries~\cite{amjad2021trackersift}.

Ultimately, the best defense will be a combination of the various defense
categories. If browsers block known tracking scripts and certain APIs used
primarily for tracking (the blocking category), limit the information provided
by other APIs (reducing entropy), and spoof values for high-entropy attributes
without producing discrepancies (increasing entropy), then it will be much
harder for fingerprinting libraries to track users and produce small anonymity
sets. These efforts, combined with users downloading extensions that block
trackers, can help limit fingerprinting on the web.

\subsection{Future Work}

One area of future work is to understand concretely how fingerprints get used in
the ad ecosystem. While we know they are used to track user interests, no one
has identified exactly which fingerprinting libraries sell information to which
advertisers. Prior work has identified the relationships between general
trackers and advertisers, and this methodology can be applied in the context of
fingerprinting~\cite{musa2022atom}. Understanding these relationships could also
uncover if a website only fingerprints users for anti-abuse purposes; if
fingerprinting is detected on a website but does not affect advertising, then we
could trust websites that claim fingerprinting is used for only anti-abuse.

As mentioned in Sections~\ref{sec:adoption} and~\ref{sec:antiabuse:adoption},
there is a need for more targeted measurement studies to understand how
fingerprinting works beyond general adoption studies, especially in the context
of anti-abuse and authentication. However, measuring fingerprinting will be more
challenging in the future because of increasing obfuscation.

As browsers and extensions focus more attention on fingerprinting defenses,
fingerprinting libraries will start to obfuscate their code more heavily. This
poses a challenge for any future measurement studies. While dynamic analysis can
get around some obfuscation, Iqbal et al. note that libraries can also try to
split their API calls across multiple scripts~\cite{iqbal2021fingerprinting}.
So, each individual script would look benign, even if it is clear that the
website is performing fingerprinting. Future work must either attribute
fingerprinting to the website, or develop ways to connect multiple scripts to
the same library.






\bibliographystyle{plain}
\bibliography{refs}

\begin{thebibliography}{10}

\bibitem{acar2020no}
Gunes Acar, Steven Englehardt, and Arvind Narayanan.
\newblock No boundaries: data exfiltration by third parties embedded on web
  pages.
\newblock {\em Proceedings on Privacy Enhancing Technologies},
  2020(4):220--238, 2020.

\bibitem{acar2014web}
Gunes Acar, Christian Eubank, Steven Englehardt, Marc Juarez, Arvind Narayanan,
  and Claudia Diaz.
\newblock The web never forgets: Persistent tracking mechanisms in the wild.
\newblock In {\em Proceedings of the 2014 ACM SIGSAC Conference on Computer and
  Communications Security}, pages 674--689, 2014.

\bibitem{acar2013fpdetective}
Gunes Acar, Marc Juarez, Nick Nikiforakis, Claudia Diaz, Seda G{\"u}rses, Frank
  Piessens, and Bart Preneel.
\newblock Fpdetective: dusting the web for fingerprinters.
\newblock In {\em Proceedings of the 2013 ACM SIGSAC conference on Computer \&
  communications security}, pages 1129--1140, 2013.

\bibitem{al2020too}
Nasser~Mohammed Al-Fannah and Chris Mitchell.
\newblock Too little too late: can we control browser fingerprinting?
\newblock {\em Journal of Intellectual Capital}, 2020.

\bibitem{alaca2016device}
Furkan Alaca and Paul~C Van~Oorschot.
\newblock Device fingerprinting for augmenting web authentication:
  classification and analysis of methods.
\newblock In {\em Proceedings of the 32nd annual conference on computer
  security applications}, pages 289--301, 2016.

\bibitem{amjad2021trackersift}
Abdul~Haddi Amjad, Danial Saleem, Muhammad~Ali Gulzar, Zubair Shafiq, and
  Fareed Zaffar.
\newblock Trackersift: Untangling mixed tracking and functional web resources.
\newblock In {\em Proceedings of the 21st ACM Internet Measurement Conference},
  pages 569--576, 2021.

\bibitem{andriamilanto2021large}
Nampoina Andriamilanto, Tristan Allard, Ga{\"e}tan Le~Guelvouit, and Alexandre
  Garel.
\newblock A large-scale empirical analysis of browser fingerprints properties
  for web authentication.
\newblock {\em ACM Transactions on the Web (TWEB)}, 16(1):1--62, 2021.

\bibitem{ayenson2011flash}
Mika~D Ayenson, Dietrich~James Wambach, Ashkan Soltani, Nathan Good, and
  Chris~Jay Hoofnagle.
\newblock Flash cookies and privacy ii: Now with html5 and etag respawning.
\newblock {\em Available at SSRN 1898390}, 2011.

\bibitem{burgess2018tyranny}
Matt Burgess.
\newblock The tyranny of gdpr popups and the websites failing to adapt.
\newblock {\em Wired}, 2018.

\bibitem{das2018web}
Anupam Das, Gunes Acar, Nikita Borisov, and Amogh Pradeep.
\newblock The web's sixth sense: A study of scripts accessing smartphone
  sensors.
\newblock In {\em Proceedings of the 2018 ACM SIGSAC Conference on Computer and
  Communications Security}, pages 1515--1532, 2018.

\bibitem{drakonakis2020cookie}
Kostas Drakonakis, Sotiris Ioannidis, and Jason Polakis.
\newblock The cookie hunter: Automated black-box auditing for web
  authentication and authorization flaws.
\newblock In {\em Proceedings of the 2020 ACM SIGSAC Conference on Computer and
  Communications Security}, pages 1953--1970, 2020.

\bibitem{duhigg2012companies}
Charles Duhigg.
\newblock How companies learn your secrets.
\newblock {\em The New York Times Magazine}, 2012.

\bibitem{durey2021fp}
Antonin Durey, Pierre Laperdrix, Walter Rudametkin, and Romain Rouvoy.
\newblock Fp-redemption: Studying browser fingerprinting adoption for the sake
  of web security.
\newblock In {\em International Conference on Detection of Intrusions and
  Malware, and Vulnerability Assessment}, pages 237--257. Springer, 2021.

\bibitem{eckersley2010unique}
Peter Eckersley.
\newblock How unique is your web browser?
\newblock In {\em International Symposium on Privacy Enhancing Technologies
  Symposium}, pages 1--18. Springer, 2010.

\bibitem{englehardt2020firefox}
Steven Englehardt.
\newblock Firefox 72 blocks third-party fingerprinting resources.
\newblock {\em Mozilla Security Blog}, 2020.
\newblock
  \url{https://blog.mozilla.org/security/2020/01/07/firefox-72-fingerprinting/}.

\bibitem{englehardt2016online}
Steven Englehardt and Arvind Narayanan.
\newblock Online tracking: A 1-million-site measurement and analysis.
\newblock In {\em Proceedings of the 2016 ACM SIGSAC conference on computer and
  communications security}, pages 1388--1401, 2016.

\bibitem{englehardt2015cookies}
Steven Englehardt, Dillon Reisman, Christian Eubank, Peter Zimmerman, Jonathan
  Mayer, Arvind Narayanan, and Edward~W Felten.
\newblock Cookies that give you away: The surveillance implications of web
  tracking.
\newblock In {\em Proceedings of the 24th International Conference on World
  Wide Web}, pages 289--299, 2015.

\bibitem{fietkau2021elephant}
Julian Fietkau, Kashyap Thimmaraju, Felix Kybranz, Sebastian Neef, and
  Jean-Pierre Seifert.
\newblock The elephant in the background: A quantitative approachto empower
  users against web browser fingerprinting.
\newblock In {\em Proceedings of the 20th Workshop on Workshop on Privacy in
  the Electronic Society}, pages 167--180, 2021.

\bibitem{fouad2022my}
Imane Fouad, Cristiana Santos, Arnaud Legout, and Nataliia Bielova.
\newblock My cookie is a phoenix: detection, measurement, and lawfulness of
  cookie respawning with browser fingerprinting.
\newblock In {\em PETS 2022-22nd Privacy Enhancing Technologies Symposium},
  2022.

\bibitem{goethem2016accelerometer}
Tom~Van Goethem, Wout Scheepers, Davy Preuveneers, and Wouter Joosen.
\newblock Accelerometer-based device fingerprinting for multi-factor mobile
  authentication.
\newblock In {\em International Symposium on Engineering Secure Software and
  Systems}, pages 106--121. Springer, 2016.

\bibitem{iab2022outlook}
IAB and PwC.
\newblock Outlook 2022: The us digital advertising ecosystem, 2022.

\bibitem{iqbal2021fingerprinting}
Umar Iqbal, Steven Englehardt, and Zubair Shafiq.
\newblock Fingerprinting the fingerprinters: Learning to detect browser
  fingerprinting behaviors.
\newblock In {\em 2021 IEEE Symposium on Security and Privacy (SP)}, pages
  1143--1161. IEEE, 2021.

\bibitem{laperdrix2020browser}
Pierre Laperdrix, Nataliia Bielova, Benoit Baudry, and Gildas Avoine.
\newblock Browser fingerprinting: A survey.
\newblock {\em ACM Transactions on the Web (TWEB)}, 14(2):1--33, 2020.

\bibitem{lin2022phish}
Xu~Lin, Panagiotis Ilia, Saumya Solanki, and Jason Polakis.
\newblock Phish in sheep's clothing: Exploring the authentication pitfalls of
  browser fingerprinting.
\newblock In {\em 31st USENIX Security Symposium (USENIX Security 22)}, pages
  1651--1668, 2022.

\bibitem{mayer2009any}
Jonathan~R Mayer.
\newblock ``any person... a pamphleteer'': Internet anonymity in the age of web
  2.0.
\newblock {\em Undergraduate Senior Thesis, Princeton University}, 85, 2009.

\bibitem{mills2009device}
Elinor Mills.
\newblock Device identification in online banking is privacy threat, expert
  says.
\newblock CNET, 2009.

\bibitem{mowery2012pixel}
Keaton Mowery and Hovav Shacham.
\newblock Pixel perfect: Fingerprinting canvas in html5.
\newblock {\em Proceedings of W2SP}, 2012, 2012.

\bibitem{firefox3pc}
Mozilla.
\newblock Firefox rolls out total cookie protection by default to all users
  worldwide.
\newblock Mozilla Blog, 2022.

\bibitem{musa2022atom}
Maaz~Bin Musa and Rishab Nithyanand.
\newblock Atom: Ad-network tomography.
\newblock {\em Proceedings on Privacy Enhancing Technologies}, 4:295--313,
  2022.

\bibitem{nikiforakis2015privaricator}
Nick Nikiforakis, Wouter Joosen, and Benjamin Livshits.
\newblock Privaricator: Deceiving fingerprinters with little white lies.
\newblock In {\em Proceedings of the 24th International Conference on World
  Wide Web}, pages 820--830, 2015.

\bibitem{nikiforakis2013cookieless}
Nick Nikiforakis, Alexandros Kapravelos, Wouter Joosen, Christopher Kruegel,
  Frank Piessens, and Giovanni Vigna.
\newblock Cookieless monster: Exploring the ecosystem of web-based device
  fingerprinting.
\newblock In {\em 2013 IEEE Symposium on Security and Privacy}, pages 541--555.
  IEEE, 2013.

\bibitem{olejnik2017battery}
Lukasz Olejnik, Steven Englehardt, and Arvind Narayanan.
\newblock Battery status not included: Assessing privacy in web standards.
\newblock In {\em 2017 International Workshop on Privacy Engineering}, 2017.

\bibitem{pinto2022fingerprintjs}
Dan Pinto.
\newblock Fingerprintjs is now fingerprint.
\newblock Fingerprint Blog, 2022.

\bibitem{preuveneers2015smartauth}
Davy Preuveneers and Wouter Joosen.
\newblock Smartauth: dynamic context fingerprinting for continuous user
  authentication.
\newblock In {\em Proceedings of the 30th annual ACM symposium on applied
  computing}, pages 2185--2191, 2015.

\bibitem{pugliese2020long}
Gaston Pugliese, Christian Riess, Freya Gassmann, and Zinaida Benenson.
\newblock Long-term observation on browser fingerprinting: Users' trackability
  and perspective.
\newblock {\em Proc. Priv. Enhancing Technol.}, 2020(2):558--577, 2020.

\bibitem{rudametkin2021improving}
Walter Rudametkin.
\newblock {\em Improving the Security and Privacy of the Web through Browser
  Fingerprinting}.
\newblock PhD thesis, Universit{\'e} de Lille, 2021.

\bibitem{chrome3pc}
Justin Schuh.
\newblock Building a more private web: A path towards making third party
  cookies obsolete.
\newblock Chromium Blog, 2020.

\bibitem{soltani2010flash}
Ashkan Soltani, Shannon Canty, Quentin Mayo, Lauren Thomas, and Chris~Jay
  Hoofnagle.
\newblock Flash cookies and privacy.
\newblock In {\em 2010 AAAI Spring Symposium Series}, 2010.

\bibitem{starov2017xhound}
Oleksii Starov and Nick Nikiforakis.
\newblock Xhound: Quantifying the fingerprintability of browser extensions.
\newblock In {\em 2017 IEEE Symposium on Security and Privacy (SP)}, pages
  941--956. IEEE, 2017.

\bibitem{sweeney2002achieving}
Latanya Sweeney.
\newblock Achieving k-anonymity privacy protection using generalization and
  suppression.
\newblock {\em International Journal of Uncertainty, Fuzziness and
  Knowledge-Based Systems}, 10(05):571--588, 2002.

\bibitem{uaclienthints}
Mike Taylor and Yoav Weiss.
\newblock User-agent client hints.
\newblock W3C Draft Community Group Report, September 2022.

\bibitem{torres2015fp}
Christof~Ferreira Torres, Hugo Jonker, and Sjouke Mauw.
\newblock Fp-block: usable web privacy by controlling browser fingerprinting.
\newblock In {\em European Symposium on Research in Computer Security}, pages
  3--19. Springer, 2015.

\bibitem{unger2013shpf}
Thomas Unger, Martin Mulazzani, Dominik Fr{\"u}hwirt, Markus Huber, Sebastian
  Schrittwieser, and Edgar Weippl.
\newblock Shpf: Enhancing http (s) session security with browser
  fingerprinting.
\newblock In {\em 2013 International Conference on Availability, Reliability
  and Security}, pages 255--261. IEEE, 2013.

\bibitem{weinshel2019oh}
Ben Weinshel, Miranda Wei, Mainack Mondal, Euirim Choi, Shawn Shan, Claire
  Dolin, Michelle~L Mazurek, and Blase Ur.
\newblock Oh, the places you've been! user reactions to longitudinal
  transparency about third-party web tracking and inferencing.
\newblock In {\em Proceedings of the 2019 ACM SIGSAC Conference on Computer and
  Communications Security}, pages 149--166, 2019.

\bibitem{safari3pc}
John Wilander.
\newblock Full third-party cookie blocking and more.
\newblock WebKit, 2020.

\bibitem{yuan2014survey}
Yong Yuan, Feiyue Wang, Juanjuan Li, and Rui Qin.
\newblock A survey on real time bidding advertising.
\newblock In {\em Proceedings of 2014 IEEE International Conference on Service
  Operations and Logistics, and Informatics}, pages 418--423. IEEE, 2014.

\end{thebibliography}

\end{document}